\documentclass{article}

\usepackage{float}
\usepackage{arxiv}
\usepackage{natbib}
\usepackage[utf8]{inputenc} % allow utf-8 input
\usepackage[T1]{fontenc}    % use 8-bit T1 fonts
\usepackage{hyperref}       % hyperlinks
\usepackage{url}            % simple URL typesetting
\usepackage{booktabs}       % professional-quality tables
\usepackage{amsfonts}       % blackboard math symbols
\usepackage{nicefrac}       % compact symbols for 1/2, etc.
\usepackage{microtype}      % microtypography
\usepackage{lipsum}
\usepackage{tablefootnote}
\usepackage{graphicx}
\usepackage{xcolor}
\graphicspath{ {./images/} }
\usepackage{subfig}
\usepackage{multirow}
\usepackage{amssymb}
\usepackage[table]{xcolor}
\usepackage[most]{tcolorbox}
\usepackage[dvipsnames]{xcolor}
\usepackage{subfig}
\usepackage[title]{appendix}

\definecolor{orange}{HTML}{ad7c28} 
\definecolor{lightorange}{HTML}{faebd2} 

\makeatletter
\renewcommand\@fnsymbol[1]{%
  \ifcase#1\or \dagger\or \ddagger\or \mathsection\or \mathparagraph\or
  \|\or **\or \dagger\dagger \or \ddagger\ddagger \fi}
\makeatother

\renewcommand{\today}{}

\usepackage{fancyhdr}
\fancyheadoffset{0pt}
\title{Generative Query Expansion with Multilingual LLMs for Cross-Lingual Information Retrieval}

\author{
  Olivia Macmillan-Scott\thanks{Work done during an internship at The Alan Turing Institute.}\\
  The Alan Turing Institute \\
  University College London \\
  \And
   Roksana Goworek\footnotemark[1]\\
  The Alan Turing Institute\\
  Queen Mary University of London\\
   \And
  Eda B. Özyiğit  \\
  The Alan Turing Institute\\
}

\begin{document}
\maketitle
\vspace{-3em}
\begin{center}

\texttt{\{omacmillan-scott, rgowerek, eozyigit\}@turing.ac.uk}
\end{center}
\vspace*{2em}

% \begingroup\def\thefootnote{*}\footnotetext{Work done during the internship at The Alan Turing Institute.}\endgroup
\setcounter{footnote}{0}

\begin{abstract}

Query expansion is the reformulation of a user query by adding semantically related information, and is an essential component of monolingual and cross-lingual information retrieval used to ensure that relevant documents are not missed. Recently, multilingual large language models (mLLMs) have shifted query expansion from semantic augmentation with synonyms and related words to pseudo-document generation. Pseudo-documents both introduce additional relevant terms and bridge the gap between short queries and long documents, which is particularly beneficial in dense retrieval. This study evaluates recent mLLMs and fine-tuned variants across several generative expansion strategies to identify factors that drive cross-lingual retrieval performance. Results show that query length largely determines which prompting technique is effective, and that more elaborate prompts often do not yield further gains. Substantial linguistic disparities persist: cross-lingual query expansion can produce the largest improvements for languages with the weakest baselines, yet retrieval is especially poor between languages written in different scripts. Fine-tuning is found to lead to performance gains only when the training and test data are of similar format. These outcomes underline the need for more balanced multilingual and cross-lingual training and evaluation resources.

\end{abstract}

% keywords can be removed

\section{Introduction}

Cross-lingual information retrieval (CLIR) is the task of identifying documents that are relevant to a given query in setting where the query and the documents are in different languages \cite{galuščáková_2022, goworek_2025}. Within the retrieval pipeline, query expansion is used so that relevant documents are not missed and a larger share of the relevant material is retrieved, which is captured by evaluation metrics such as recall. In this context, query expansion refers to the automatic modification of a user query by adding related terms or generated text in order to better encompass the user's information requirements.

\begin{figure}[h!]%
    \centering{\includegraphics[width=1\linewidth]{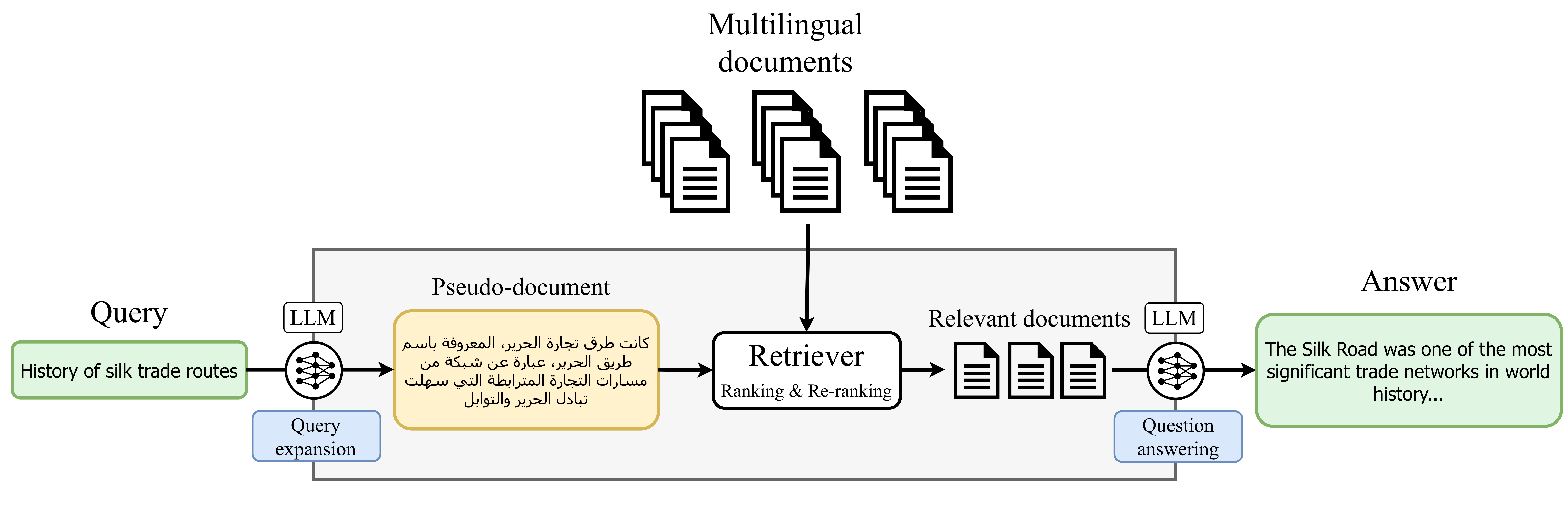}}
    \caption{Overview of an information retrieval pipeline. Query expansion (along with question answering) is one of the stages that has most benefitted from recent developments of transformer-based models}%
    \label{fig:ir}%
\end{figure}

The development of transformer-based large language models (LLMs) has enabled novel query expansion techniques that leverage the generative capabilities of these models to create pseudo-documents, which are then used as expanded queries \cite{li_2025_qe}. The retrieval pipeline, both in monolingual information retrieval (where the query and documents are in the same language) and in CLIR, typically consists of four main components: query expansion, ranking, re-ranking, and question answering \cite{zhu_2024_ir}. As illustrated in Figure~\ref{fig:ir}, the two steps that require text generation, and therefore, can arguably benefit the most from recent advances in generative models, are query expansion and question answering. Of the two components, this paper focuses on query expansion, as it is the first stage at which generative models can directly influence which documents are retrieved.

Although most work in this area has focused on monolingual retrieval, the increased multilingual capabilities of generative models make them a valuable tool for cross-lingual applications \cite{xu_2025, qin_2025}. Information retrieval (IR) is used in a wide range of applications, many of which operate in multilingual settings. For instance, IR is a central component of search engines and can be used to find online content in different languages \cite{buttcher_2010, croft_2009}. Modern systems now often include a question answering step, in which the relevant information extracted from documents is presented in the language of the original query \cite{jurafsky_2025, asai_2021_qa, muller_2022}. CLIR can therefore democratise access to information by allowing users to obtain content that is available in languages they do not understand. The potential benefits span many domains, from language agnostic access to scientific literature, to faster collection of intelligence in critical security situations. Query expansion is a key part of such retrieval systems, since it helps ensure that retrieved documents better match the user information needs and reduces the mismatch between short, ambiguous queries and longer-form documents.

This paper analyses state-of-the-art methods for query expansion based on LLMs and applies them to the cross-lingual case with the aim of establishing which approaches perform best under given conditions, ultimately improving CLIR performance. The introduction of such models has shifted query expansion from traditional semantic expansion, where a query is augmented with synonyms and related terms, to the generation of pseudo-documents that are used as expanded queries at retrieval time. The study evaluates several models and expansion techniques in order to identify which factors most strongly influence retrieval quality. The experimental setup covers a wide range of both high and low resource languages that use different scripts, making it possible to also highlight differences in performance across languages. The following four research questions (RQs) guide this work:

\textbf{RQ1: What multilingual LLMs show the best performance on cross-lingual query expansion?}  Different models differ in their training data, in the extent of their multilingual coverage, and in the degree to which they are adapted to produce text that is relevant to a given query. Comparing several multilingual models makes it possible to analyse how such factors influence cross-lingual query expansion and to examine whether particular expansion strategies are better suited to particular model families.

%We find that Gemma 3 12B performs best on almost every metric, with bigger improvements compare to Gemma 3 4B than Aya Expanse 8B.

\textbf{RQ2: How do different expansion techniques perform for varying query lengths?} The study uses two datasets with queries and documents that differ in format, source and, of particular interest, in average query length. Query length is an important factor, since the ambiguity of short queries is often a central cause of poor retrieval performance. By contrasting datasets with shorter and longer queries, it becomes possible to investigate whether specific expansion techniques are more effective for one type of query than another.

%We find that for shorter queries, zero-shot prompting performs better than more elaborate prompting techniques, whereas for longer queries we instead see few-shot prompting leading to higher retrieval performance. 

\textbf{RQ3: How do results vary across languages?} A central issue in cross-lingual retrieval is the imbalance between languages and the large differences in performance that often arise. The datasets considered here include a range of languages with different resource levels and scripts. This setting allows for a systematic examination of how linguistic properties and resource availability affect cross-lingual query expansion and the resulting retrieval quality.

%The highest retrieval performance is found for European language pairs, with the lowest performance often on Arabic or Chinese queries and documents. However, some of the languages with the lowest absolute metrics also see the biggest improvements through cross-lingual query expansion.
   
\textbf{RQ4: Does supervised fine-tuning improve query expansion performance?} Much of the existing work on generative models for query expansion has focused on monolingual, primarily English, settings and has relied heavily on prompting. This study extends current work by applying supervised fine-tuning on cross-lingual retrieval data and by comparing the behaviour of the fine-tuned model on original queries and on expanded queries. This setting also makes it possible to explore how prompting and fine-tuning interact when used jointly for cross-lingual query expansion.

%We do find improves retrieval performance when using the fine-tuned models for cross-lingual query expansion, and CoT prompting now appears as the best-performing approach.
%\end{itemize}

We find that different query expansion approaches lead to better retrieval when used with certain models and query types, where shorter queries benefit from less elaborate prompting techniques. Higher performance is found for Latin-script, European language pairs, and fine-tuning results in performance gains only when the training and test data are of similar format. The contributions of this work are threefold: (i) it provides a comprehensive comparative analysis of generative query expansion techniques using multilingual large language models; (ii) it shows that different approaches are better suited to specific query formats, so that no single method consistently dominates; and (iii) it extends existing work by combining established query expansion techniques with multilingual fine-tuning, yielding improvements in retrieval performance for similar queries, and losses for different-style queries.

The remainder of this paper is structured as follows: Section \ref{sec:related} reviews relevant literature and developments in query expansion for information retrieval. Section \ref{sec:methods} describes the methodology, including the datasets and models, as well as the experimental setup and evaluation metrics. Section \ref{sec:results} presents and discusses the results obtained for the various models and expansion approaches, both in aggregate and at the level of individual datasets. Lastly, Section \ref{sec:conclusion} concludes the paper and outlines limitations and directions for future work.

\section{Related Work}
\label{sec:related}

This section reviews related work on cross-lingual information retrieval and its recent evolution from lexical ranking to embedding-based neural methods, with a particular focus on multilingual and cross-lingual settings. It then surveys query expansion techniques, tracing developments from traditional term-based approaches to LLM-based methods that use prompting, fine-tuning and different combinations of translation and expansion. Finally, it discusses recent multilingual LLMs designed for generative query expansion, highlighting their architectural properties and multilingual capabilities that are most relevant for cross-lingual retrieval.

\subsection{Information Retrieval} 

In recent years, CLIR research has benefitted substantially from advances in both monolingual information retrieval and multilingual natural language processing. Widely used ranking algorithms based on lexical matching, such as TF-IDF~\cite{sparck_1988} and BM25~\cite{robertson_2009}, have increasingly been complemented or replaced by embedding-based neural methods that compute vector similarities between encoded queries and documents \cite{shi_2021, huang_2024, weller_2025, feng_2022, reimers_2020}. Dense retrieval approaches typically employ bi-encoder architectures for the initial retrieval stage and cross-encoder architectures for re-ranking \cite{goworek_2025,xu_2025_ir}. Transformer-based models have further enabled retrieval across languages without the need for intermediate translation. Several recent large language models have been explicitly trained for multilingual capabilities (for surveys, see \cite{, xu_2025, qin_2024, zhu_2024_mllms, gurgurov_2024, huang_2025}). The overarching objective is to obtain language agnostic representations that support efficient search in multilingual settings. Despite these advances, significant disparities remain across languages, with markedly reduced performance for lower-resource and typologically distant languages \cite{galuščáková_2022}. A small number of surveys encompass both the cross-lingual and retrieval aspects of CLIR while also including recent embedding-based neural approaches \cite{galuščáková_2022, goworek_2025}, whereas earlier overviews of CLIR focus mainly on translation-based or lexical methods and do not consider dense retrieval  \cite{nie_2010}. Related surveys on monolingual retrieval also provide useful background for understanding these developments \cite{zhu_2024_ir, xu_2025_ir,hambarde_2023}.

\subsection{Query Expansion} 

In both monolingual and cross-lingual retrieval, query expansion aims to address two main issues. First is the \textit{vocabulary problem} \cite{furnas_1987}, which refers to the mismatch between the terms used in the query and those used to index the documents, resulting in low recall as relevant documents are missed. Second, the query may not accurately reflect the user's information requirements, again leading to poor recall. The short length of queries, on average less than 2.5 words long for web search \cite{jansen_2000, spink_2001}, means that they contain significant ambiguity and may not provide sufficient context for effective retrieval. Several factors exacerbate the vocabulary problem and the ambiguity of queries, including polysemy (where one word may contain multiple meanings), word inflections, and the presence of multiple potential translation or transliterations in the cross-lingual case. Query expansion is a useful tool to mitigate some of these issues, and has been found to be particularly effective to address poor retrieval from ambiguous and lower-quality queries. 

Query expansion initially involved the addition of relevant terms or phrases to the query, with this augmented query then being used in the retrieval process. The terms used in the expansion could come from either \textit{global} data, leveraging external resources such as thesauri or ontologies, or \textit{local} data, instead relying on the document set that the retrieval will be performed on \cite{xu_croft_2017}. The latter was pioneered by Rocchio \cite{rocchio_1965, rocchio_1971}, who used user-identified relevant documents in a first retrieval pass to identify expansion terms. This method, known as relevance feedback, was then followed by pseudo-relevance feedback (PRF) \cite{croft_harper_1979}, which employs a similar approach but eliminates the user's input by assuming the top-retrieved documents are relevant. RM3 is a widely used model for both relevance feedback and PRF \cite{lavrenko_2001}.

%Traditional QE approaches can be categorised into those using global data, leveraging external resources like thesauri or ontologies, and those using local data, which take expansion terms from initially retrieved documents \citep{xu_croft_2017}.

Neural network-based language models have allowed for the development of new query expansion techniques that do not need to rely on hand-built resources or relevance feedback, and in which the vocabulary mismatch problem is at least partially addressed through the use of continuous embeddings \cite{li_2025_qe, yao_2025, yang_2025}. Instead of augmenting the original query with additional related terms and phrases, recent methods leverage the text generation capabilities of LLMs to generate pseudo-documents in response to the query, and use this in the retrieval process. This approach is particularly beneficial for embedding-based retrieval, where the difference between short-form queries and long-form documents can lead to poor retrieval performance. For monolingual retrieval, three main LLM-based approaches to query expansion have been identified: prompting, supervised fine-tuning, and reinforcement learning \cite{li_2025_qe, zhu_2024_ir}. 

The majority of current research has focused on prompting approaches for query expansion; not only are they the simplest to implement, but they also require no additional resources or training data \cite{wang_2023_query2doc, gao_2023_hyde, shen_2024_lamer, liu_2025_exp4fuse}. For instance, building on relevance feedback and PRF, some have proposed Generative Relevance Feedback (GRF) and have shown that it can significantly outperform PRF approaches \cite{mackie_2023, claveau_2022}. In such settings, it is no longer necessary to select and rank individual expansion terms explicitly. Supervised fine-tuning, in contrast, requires the construction of datasets specifically designed for query expansion \cite{zhu_2024_ir}. Some have also looked at the use of reinforcement learning for query expansion \cite{zhao_2025}, although this represents a minority of existing work in the area.

A particularly influential prompting paradigm, often referred to as answer-incorporated queries, is the generation of a pseudo-document from the initial query to form an expanded representation, sometimes concatenating the generated text with the original query. This approach has been shown to significantly improve retrieval performance, and has seen varying implementations for both sparse and dense retrieval. Hypothetical Document Embeddings (HyDE)~\cite{gao_2023_hyde}, Multi-Text Generation Integration (MuGI)~\cite{zhang_2024_mugi} and Exp4Fuse \cite{liu_2025_exp4fuse} use zero-shot prompting, whereas others explore more elaborate prompting techniques \cite{shen_2024_lamer, jagerman_2023, deng_2024_rar}, including the use of few-shot prompting in Query2doc~\cite{wang_2023_query2doc}. However, some have also pointed out the limitations of using LLM-based query expansion, primarily relating to the introduction of irrelevant information due to either a lack of domain knowledge or ambiguity in the original query \cite{abe_2025}.

Much of this research has been conducted in monolingual settings, yet cross-linguality introduces an additional layer of complexity. As mentioned, queries used for information retrieval are typically short and therefore ambiguous, which can lead to translation errors. Three main strategies have been explored in the cross-lingual literature: pre-translation expansion, post-translation expansion, and a combination of the two. Applying translation before expansion has often been found to be the most successful strategy overall \cite{ballesteros_croft_1998}, although there is a lack of recent work comparing the two. Pre-translation expansion tends to improve precision by refining the query before it is applied to the target language, whereas recall is typically enhanced by post-translation expansion, which augments the query after translation and can therefore capture a broader range of relevant documents. In this work, we evaluate both orders of implementation; Figure \ref{eg:t+e} illustrates an example where the query is first translated and then expanded.

\begin{figure}[h!]
\begin{tcolorbox}[colback=lightorange,colframe=orange,boxsep=2pt,arc=1pt, title={Zero-Shot Query Expansion for mMARCO: French $\rightarrow$ English (translation + expansion)}]
\textbf{Query:} nom d'origine du mercure

\medskip

\textit{Query is translated from French to English}

\medskip

\textbf{Translated query:} original name of mercury

\medskip

\textbf{LLM Prompt:} 

Please write a passage to answer the question.\\
Question: original name of mercury

\medskip

\textit{Query is expanded via zero-shot prompting}

\medskip

\textbf{Pseudo-document:} 

The name ``Mercury'' originates from Roman mythology. It is derived from the Roman god Mercury (or Mercurius in Latin), the equivalent of the Greek god Hermes. Mercury was considered the messenger of the gods, known for his speed and wit. In the solar system, the planet closest to the Sun has been named in his honor, reflecting its swift motion and proximity to our star.
\end{tcolorbox}
\caption{Zero-shot query expansion example for mMARCO dataset \cite{bonifacio_2022_mmarco}.}
\label{eg:t+e}
\end{figure}

\subsection{Multilingual LLMs (mLLMs) for Query Expansion}

Building on the use of LLMs for query expansion in monolingual retrieval, mLLMs can be used in a similar way to expand queries in cross-lingual settings, mirroring the shift from lexical and semantic expansion to the generation of pseudo-documents. Within this context, specific open weight architectures have emerged as important tools for generative query expansion. In this work, three open-source mLLMs are evaluated. Cohere for AI's Aya Expanse 8B~\cite{dang_2024_aya} is an 8 billion parameter mLLM that is specifically optimised for high performance across 23 languages. Its multilingual capabilities are supported by training techniques such as data arbitrage and multilingual preference training, which enable the model to approach the performance of substantially larger monolingual systems. The Gemma 3 family~\cite{gemma_2025} developed by Google DeepMind, in particular the 4B and 12B variants, represents a recent generation of compact open models derived from the same underlying technology as Gemini. These models introduce features that are beneficial for generative query expansion in cross-lingual retrieval, including multilingual support for more than 140 languages and improved efficiency at inference time. Taken together, these architectures illustrate rapid progress toward powerful, efficient, and broadly multilingual models that can streamline the joint translation and expansion step required for effective retrieval across languages.

\section{Experimental Setup}
\label{sec:methods}

\subsection{Overview}

This study distils the main components of existing generative query expansion techniques according to prompting method and what forms the expanded query in order to enable a systematic comparison of their effectiveness. Retrieval performance is evaluated for four prompting strategies: zero-shot prompting, Chain-of-Thought (CoT) prompting,  Rephrase and Respond (RaR), and  few-shot prompting. The exact prompts used are listed in Appendix \ref{app:prompts}. In addition, the experimental design builds on these state-of-the-art methods by using using fine-tuning along with the aforementioned prompting techniques: two fine-tuned variants of a base mLLM are created by using different language subsets of CLIR data, and the performance of the base model and fine-tuned variants are compared. For each approach, retrieval is carried out both using only the pseudo-document and using a concatenation of the original query and the pseudo-document.

Much of the existing research has been conducted in monolingual settings, yet cross-linguality introduces an additional layer of complexity to query expansion. Queries used for information retrieval are typically short and therefore ambiguous, which can lead to translation errors. Ballesteros and Croft \cite{ballesteros_croft_1998} reported higher recall when translating and then expanding the query, rather than expanding before translation. To investigate these questions, four parameters are varied in the query expansion process: (i) three open source multilingual models and two fine-tuned variants; (ii) the choice of prompting technique; (iii) the order of translation and expansion; and (iv) the use of only the pseudo-document versus the combination of pseudo-document and original query for retrieval. These configurations are summarised in Table \ref{tab:qe}, and all parameter combinations are evaluated on both CLIRMatrix and mMARCO.

\begin{table*}[h]
    \centering
    \begin{tabular}{cccc} \toprule
           \textbf{Model} &\textbf{Method} & \textbf{Order} & \textbf{Used for retrieval} \\ \midrule
           Aya Expanse 8B & Zero-shot & T + E & Doc only \\
           Gemma 3 4B & CoT & E + T & Q + doc \\
           Gemma 3 12B & RaR &  & \\
           Aya fine-tuned on CLIRMatrix En \& Es & Few-shot &  & \\ 
           Aya fine-tuned on CLIRMatrix Ar \& Zh & & & \\
         \bottomrule
    \end{tabular}
    \caption{Parameters used for query expansion; all combinations are evaluated on both CLIRMatrix and mMARCO. \\ T: translation; E: expansion.}
    \label{tab:qe}
\end{table*}

\subsection{Datasets}

Effective training and evaluation of CLIR systems require large scale datasets that cover many languages and domains. The limited availability of such resources remains a central challenge, particularly for low-resource languages and specialised applications. A substantial proportion of existing work relies on Wikipedia to construct the document corpus for retrieval \cite{goworek_2025}. Wikipedia is a valuable resource as it contains large amounts of human-generated, generally high-quality text spanning a multitude of topics. However, the reliance on a single resource reduces heterogeneity in the types of data used to train and evaluate retrieval systems. Language coverage is also restricted, with most datasets including fewer than 15 languages.

Available resources range from open-domain to domain-specific datasets, and from collections that cover typologically diverse languages to those with a regional focus. More general datasets with broad language coverage include CLIRMatrix~\cite{sun_2020_clirmatrix}, Large-scale CLIR Dataset \cite{sasaki_2018}, SWIM-IR~\cite{thakur_2024_swimir}, XOR-TyDi QA \cite{asai_2021_qa}, LAReQA~\cite{roy_2020_lareqa} and mMARCO~\cite{bonifacio_2022_mmarco}. In contrast, AfriCLIRMatrix~\cite{ogundepo_2022_africlirmatrix} and CIRAL~\cite{adeyemi_2024_ciral} are examples of datasets with a regional focus: both contain text in a range of African languages. BordIRlines~\cite{li_2024_bordirlines} is an example of a domain-specific resource, with queries and documents related to geopolitical border disputes. Further datasets have been developed primarily for question answering, again often drawing on Wikipedia, including XQA~\cite{liu_2019_xqa}, MLQA~\cite{lewis_2020_mlqa}, XQuAD~\cite{artetxe_2020}, TyDi QA~\cite{clark_2020_tydiqa}, MKQA \cite{longpre_2021_mkqa}, and XRAG \cite{liu_2025_xrag}. 

To mitigate data scarcity and increase language coverage, the use of machine-generated text and translations has grown in recent years. Machine translation has long been employed to address imbalances across languages; for instance, mMARCO \cite{bonifacio_2022_mmarco} is a machine-translated version of the English MS MARCO passage ranking dataset \cite{bajaj_2018_msmarco}. More recently, LLMs have been used to generate queries from documents and vice versa, or even to produce triplets of queries, positive passages and negative passages. This trend, however, raises concerns about noise and bias introduced by machine-generated content and about the extent to which such corpora faithfully reflect real user behaviour. 

\begin{table*}[t]
    \centering
    \begin{tabular}{ccccccccccccccccc}
    \toprule
        \textbf{Dataset} &  \textbf{\# Docs} &  \textbf{\# Queries} & Ar & De & En & Es & Fr & Hi & Id & It & Ja & Nl & Pt & Ru & Vi & Zh\\
         \midrule
         CLIRMatrix & $\thicksim$15M & 104k % 12,000 per lang. 
         & \checkmark & \checkmark & \checkmark & \checkmark & \checkmark &  &  &  & \checkmark & & & \checkmark & & \checkmark \\
         mMARCO & 123.2M & $\thicksim$7.4M & \checkmark & \checkmark & \checkmark & \checkmark & \checkmark & \checkmark & \checkmark & \checkmark & \checkmark & \checkmark & \checkmark & \checkmark & \checkmark & \checkmark \\
         \bottomrule
    \end{tabular}
    \caption{Composition of datasets used; number of documents and queries are the total for all languages (e.g. mMARCO contains 8.8M documents and 530k queries translated from English into each language).}
    \label{tab:data}
\end{table*}

This paper uses CLIRMatrix~\cite{sun_2020_clirmatrix} and mMARCO~\cite{bonifacio_2022_mmarco}], which together provide a set of typologically diverse languages (see Table \ref{tab:data}; for language codes, see Table \ref{tab:langs}), and differ substantially in data characteristics. CLIRMatrix contains Wikipedia articles and derived queries, whereas mMARCO consists of translated question answering style queries paired with short passages.

% \begin{table*}
%     \centering
%     \begin{tabular}{llll}
%     \toprule
%     \textbf{ISO Code} & \textbf{Language}  & \textbf{ISO Code} &\textbf{Language}  \\
%     \midrule
%  Ar & Arabic  & It &Italian  \\
%  De & German  & Ja &Japanese  \\
%  En & English  & Nl &Dutch  \\
%  Es & Spanish  & Pt &Portuguese  \\
%  Fr & French  & Ru &Russian  \\
%  Hi & Hindi  & Vi &Vietnamese  \\
%  Id & Indonesian  & Zh &Chinese  \\
%  \bottomrule
%     \end{tabular}
%     \caption{Language ISO codes and corresponding languages.}
%     \label{tab:langs}
% \end{table*}

\begin{table*}
    \centering
    \begin{tabular}{lllll}
    \cmidrule[\heavyrulewidth](lr){1-2}
    \cmidrule[\heavyrulewidth](lr){4-5}

    \textbf{Code} & \textbf{Language} && \textbf{Code} & \textbf{Language} \\
    \cmidrule(lr){1-2}
    \cmidrule(lr){4-5}

    Ar & Arabic      && It & Italian     \\
    De & German      && Ja & Japanese    \\
    En & English     && Nl & Dutch       \\
    Es & Spanish     && Pt & Portuguese  \\
    Fr & French      && Ru & Russian     \\
    Hi & Hindi       && Vi & Vietnamese  \\
    Id & Indonesian  && Zh & Chinese     \\

    \cmidrule[\heavyrulewidth](lr){1-2}
    \cmidrule[\heavyrulewidth](lr){4-5}
    \end{tabular}

    \caption{ISO codes and corresponding languages.}
    \label{tab:langs}
\end{table*}

\paragraph{CLIRMatrix.}

CLIRMatrix \footnote{https://www.cs.jhu.edu/\string~shuosun/clirmatrix} \cite{sun_2020_clirmatrix} comprises two datasets: a bilingual collection in 139 languages, with queries in one language and documents in another (19,182 language pairs in total), and a multilingual collection in eight languages, in which queries and documents are jointly aligned. The experiments reported here use the multilingual dataset, which contains queries and documents in Arabic, German, English, Spanish, French, Japanese, Russian and Chinese (see Table \ref{tab:data}), as this allows for cross-lingual retrieval across many aligned language pairs. 

In CLIRMatrix, the Wikipedia article titles form the dataset's queries. Sun and Duh~\cite{sun_2020_clirmatrix} considered using human-generated search queries but noted that such data are not publicly available for most languages. Using the title as the query can approximate short web search queries, which are often only two or three words long \cite{jansen_2000, spink_2001}. This design choice has limitations, including limited variation in query formulations and possible ambiguity in titles. Relevance judgements between queries and the Wikipedia articles that form the document corpus are obtained by applying BM25 and smoothing the resulting scores to produce discrete relevance levels. For each query, the top 100 documents are assigned relevance scores from 1 to 5 (with 5 indicating highest relevance), and the document associated with the query receives a score of 6.

\paragraph{mMARCO.} mMARCO \footnote{https://huggingface.co/datasets/unicamp-dl/mmarco} \cite{bonifacio_2022_mmarco} is a machine-translated version of the English MS MARCO passage ranking dataset \cite{bajaj_2018_msmarco}. The original dataset contains queries sampled from Bing search logs, each paired with one positive (relevant) and one negative (non-relevant) passage. For mMARCO, two translation pipelines are employed: neural translation models developed by the Language Technology Research Group at the University of Helsinki \cite{tiedemann_2020_opus} and Google Translate \footnote{https://cloud.google.com/translate}. 

The original MS MARCO dataset is translated into 13 languages, so the full dataset is made up of the 14 languages
displayed in Table \ref{tab:data}: English, Chinese, French, German, Indonesian, Italian, Portuguese, Russian, Spanish, Arabic, Dutch, Hindi, Japanese and Vietnamese. Because the original passage ranking dataset was derived from a question answering corpus, mMARCO contains relatively long queries phrased as natural language questions (e.g. ``what fruit is native to Australia") and comparatively short passages. Unlike CLIRMatrix, which provides a ranked list of documents with graded relevance labels, mMARCO offers, for each query, a single positive passage that answers the query and a hard negative passage that is topically related but does not constitute a correct answer.

% \begin{figure}[h!]%
%     \centering{\includegraphics[width=1\linewidth]{figures/mm.png}}
%     \caption{Positive and negative document examples for mMARCO dataset.}%
%     \label{fig:mm}%
% \end{figure}

% \paragraph{Large-Scale CLIR Dataset.}

% The Large-Scale CLIR Dataset \footnote{https://www.cs.jhu.edu/~kevinduh/a/wikiclir2018/} \cite{sasaki_2018} is also derived from Wikipedia articles. This dataset contains English queries with relevant document in 25 other languages. In this case, instead of using titles, the first sentence of the English Wikipedia article is extracted to act as the query. The document is the first 200 words of the article in the relevant languages. This dataset covers a wider range of languages, covering both high-resource languages such as Chinese and French, as well as low-resource languages including as Swahili and Tagalog.

\subsection{Implementation Details}

This study analyses the performance of mLLMs on pseudo-document generation for information retrieval. Three base open-source models with strong multilingual capabilities are considered: Aya Expanse 8B\footnote{https://huggingface.co/CohereLabs/aya-expanse-8b} and Gemma 3 4B\footnote{https://huggingface.co/google/gemma-3-4b-it} \& 12B\footnote{https://huggingface.co/google/gemma-3-12b-it}. Building on existing work, fine-tuned variants of Aya Expanse 8B on CLIRMatrix data are also evaluated. One variant is fine-tuned on English and Spanish, two high resource languages that share the same script and typically achieve strong results in cross-lingual retrieval. The other is fine-tuned on Arabic and Chinese, languages written in different scripts that often rank among the lowest performing in cross-lingual retrieval and other multilingual benchmarks. Comparison of these variants makes it possible to assess both the overall effect of fine-tuning on retrieval performance and the influence of the specific languages used during fine-tuning.

The query expansion approaches considered here involve both a translation and an expansion step, meaning that we carry out the two stages sequentially rather than directly going from the source-language query to the target-language pseudo-document. As a result, the isolated expansion step is applied in a monolingual setting. To support this design, the multilingual CLIRMatrix dataset is used to construct monolingual subsets for Arabic, Chinese, English and Spanish, which are then combined to create two datasets: one English and Spanish, and one Arabic and Chinese. The base Aya Expanse 8B model is then fine-tuned on each of the two datasets. These fine-tuned variants are employed for the query expansion stage in the implementation, while maintaining the original base model for the translation step.

Most existing LLM-based query expansion techniques rely on prompting. Methods such as HyDE~\cite{gao_2023_hyde}, Exp4Fuse~\cite{liu_2025_exp4fuse} and MuGI~\cite{zhang_2024_mugi} use zero-shot prompting to generate a hypothetical document that is then used for sparse or dense retrieval. Other studies compare prompting strategies, including zero-shot, few-shot and CoT prompting, and report strong performance for the latter \cite{jagerman_2023}. Deng et al. \cite{deng_2024_rar} propose the Rephrase and Respond approach for question answering, in which the model first rewrites the query to better express the underlying information need and then returns an answer; in this work, their strategy is adapted to cross-lingual retrieval by using the rephrased query as the expanded representation. Few-shot prompting has also been explored for query expansion, such as with query2doc~\cite{wang_2023_query2doc}, where example query-document pairs are first randomly sampled from a training set and presented to the model before prompting the model to generate a pseudo-document, and in LameR~\cite{shen_2024_lamer}, which incorporates relevance feedback by retrieving top-ranked documents for the input query and using them to enrich the prompt. There is substantial methodological overlap among some of these approaches; for instance, HyDE and Exp4Fuse both generate hypothetical documents that serve as expanded queries, and their main components have been distilled in this work.

\subsection{Evaluation Metrics}

There are no established intrinsic metrics that evaluate the query expansion component of the retrieval pipeline in isolation. In practice, query expansion is assessed extrinsically by using the expanded queries for retrieval and comparing the performance obtained with the original and expanded queries through standard information retrieval metrics. A baseline is first established using the original queries, and the different expansion techniques are then evaluated against this baseline. Within information retrieval, the focus is primarily on recall rather than precision, since it is generally more important that relevant documents are not missed rather than that all retrieved documents are relevant. Retrieval performance is measured using the following metrics, where $k$ denotes the number of retrieved documents (that is, evaluation is restricted to the top $k$ results): 

\paragraph{Hit@k.} Hit@k indicates whether at least one relevant document appears within the top $k$ results. Hit@1 corresponds to accuracy, as it reports whether the top ranked document is relevant.

\paragraph{Recall@k.} Recall@k measures the proportion of relevant documents retrieved in the top $k$ results out of all relevant documents for a query in the collection. This metric focuses on completeness and captures how much of the relevant information the system is able to recover. It is widely used in the information retrieval literature, particularly in settings where coverage of relevant material is critical.

\paragraph{MRR.} Mean Reciprocal Rank is the mean, across queries, of the reciprocal of the rank of the first relevant document. This metric is particularly informative for single answer tasks such as question answering and is order aware, since it explicitly reflects how early a relevant result appears in the ranking.

\paragraph{nDCG@k.}Normalized Discounted Cumulative Gain at $k$ is a graded relevance metric that takes into account both the relevance level of retrieved documents and their rank positions. Higher weight is assigned to relevant documents that appear earlier in the ranked list, and the score is normalised by the best possible ranking for each query.

% \paragraph{MAP.} Mean Average Precision reflects overall ranking quality. For each query, average precision is computed as the mean of the precision values at all ranks where a relevant document is retrieved. MAP is then obtained by averaging these values across queries. As an order aware metric, it rewards systems that retrieve relevant documents early and with few non relevant documents interspersed.

For mMARCO, only a single relevant passage is provided for each query. In this case, Recall@k and Hit@k are equivalent, since both reduce to an indicator of whether the unique relevant passage appears in the top $k$ results. Consequently, Hit@k is not reported for mMARCO in order to avoid redundant metrics. For a single relevant document, nDCG@k provides essentially the same information as MRR, so it is also not reported for mMARCO.

% (Compare to results in CLIRMatrix paper - they use BM25 (normalised?))

\section{Results and Discussion}
\label{sec:results}

This section is organised around the four research questions introduced above, with the corresponding experimental results presented in separate subsections. Section \ref{sec:comp} reports results for all three multilingual models. Subsequent subsections concentrate on Aya Expanse 8B and its fine-tuned variants, since similar patterns are observed across models. Additional results for Gemma 3 4B and 12B are provided in Appendices \ref{app:g4b} and \ref{app:g12b}.

%\subsection{RQ1: What expansion techniques and models results in the best retrieval performance?}
\subsection{RQ1: What multilingual LLMs show the best performance on cross-lingual query expansion?}
\label{sec:comp}

The first set of experiments compares the retrieval performance of the three multilingual LLMs under different cross-lingual query expansion techniques, the aggregated results of which are summarised in Table \ref{tab:summary}. For this comparison, scores are averaged across all language pairs, implementation orders, and retrieval variants (pseudo-document only versus query plus pseudo-document). Across both datasets and all metrics, Gemma 3 4B achieves the lowest performance. On mMARCO, expansion with Gemma 3 12B outperforms Aya Expanse 8B, whereas on CLIRMatrix the two models exhibit comparable behaviour, with each obtaining the best score on some metrics.

\begin{table*}[h!]
    \centering
    \begin{tabular}{ll|ccc|ccc}
\toprule
\multirow{2}{*}{\textbf{Model}} & \multirow{2}{*}{\textbf{Expansion}} &  \multicolumn{3}{c|}{\textbf{CLIRMatrix}} &  \multicolumn{3}{c}{\textbf{mMARCO}} \\
         &  &  Hit@10&  Recall@50&MRR &  Recall@10&  Recall@50&MRR\\
\hline
\rowcolor{gray!15}\multicolumn{2}{c|}{Original query} & 69.35 & 10.09 & 41.35  &38.55& 79.27& 15.39\\
\hline
\multirow{4}{*}{Aya Expanse 8B}  & Zero-shot& \underline{84.06} & \underline{\textbf{13.32}} & \underline{60.78} &  39.18&   78.69 & 18.57\\
         &  CoT& 83.38 & 13.23 & 58.93  &  40.02 &   79.13 & 19.19 \\
         &  RaR& 82.01 & 13.30 & 57.25  &  40.43 &   79.49 &19.44\\
         &  Few-shot& 81.98 & 13.29 & 57.33  &\underline{40.47} &   \underline{79.54} & \underline{19.49} \\
\hline
 \multirow{4}{*}{Gemma 3 4B} & Zero-shot& 76.98	& 11.68	& 51.03	& 39.03 &  78.23 & 18.86\\
 & CoT& \underline{77.64} &	\underline{11.80} &	\underline{51.64} & \underline{39.25} &  \underline{78.52} & \underline{19.13} \\
 & RaR& 72.37 &	10.83 &	46.36 & 38.39&  78.02 &18.63\\
 & Few-shot& 72.37 & 10.80 & 46.40 & 38.40&  78.03&18.66\\
\hline
\multirow{4}{*}{Gemma 3 12B} & Zero-shot& \underline{\textbf{85.12}} & \underline{13.24} & \underline{\textbf{62.03}} & 40.64 & 79.42 & 19.91\\
 & CoT& 82.70 & 12.97 & 59.43  & 41.65 & 80.31 & \underline{\textbf{20.38}}\\
 & RaR& 79.59 & 12.70 & 55.08  & 41.67 & 80.76 & 20.22\\
 & Few-shot& 79.59 & 12.69 & 55.11  & \underline{\textbf{41.70}} & \underline{\textbf{80.77}} & 20.26 \\
\hline
% \multirow{4}{*}{\parbox{3cm}{Aya fine-tuned  on\\ CLIRMatrix En \& Es}} & Zero-shot& &  && & &  \\
%  & CoT& &  && & &  \\
%  & RaR& &  && & &  \\
%  & Few-shot& &  && & &  \\
% \hline
% \multirow{4}{*}{\parbox{3cm}{Aya fine-tuned  on\\ CLIRMatrix Ar \& Zh}} & Zero-shot& &  && & &  \\
%  & CoT& &  && & &  \\
%  & RaR& &  && & &\\
%  & Few-shot& &  && & & \\
% \hline
\end{tabular}
    \caption{Summary of cross-lingual query expansion results across all language pairs. Note that CLIRMatrix contains 100 relevant documents per query whereas mMARCO contains one per query (so Hit@k = Recall@k). (\textbf{max value}, \underline{max value per model})}
    \label{tab:summary}
\end{table*}

The optimal prompting strategy varies across models and datasets. For Gemma 3 4B, CoT prompting consistently yields the strongest results on both datasets for all metrics. In contrast, Aya Expanse 8B and Gemma 3 12B show a clear difference between datasets: both indicate that zero-shot prompting is most effective on CLIRMatrix, while few-shot prompting leads to the best performance on mMARCO. The only exception is MRR on mMARCO for Gemma 3 12B, where CoT prompting slightly surpasses few-shot prompting. For these two larger models on mMARCO, zero-shot expansion produces the weakest retrieval performance. A central distinction between CLIRMatrix and mMARCO, which likely accounts for part of this behaviour, is the length of the queries and their relation to document length; this issue is examined in more detail in Section \ref{sec:q_len}. 

The highest retrieval metrics are therefore achieved through performing cross-lingual expansion with the multilingual LLM that has been trained on the largest number of parameters that we tested: Gemma 3 12B. This is particularly marked for mMARCO, where Gemma 3 12B outperforms the other two models in every configuration. However, for CLIRMatrix we do have some cases where Aya Expanse 8B leads to better recall and ranking scores. Some of this disparity may come from the language training: whereas Aya Expanse is optimised across 23 languages, Gemma 3 benefits from multilingual training in a broader number of languages. Both the increased number of languages in its training data and the larger number of parameters used likely contribute to the higher performance derived from the query expansion with Gemma 3 12B. Having said that, Aya Expanse's pre-training data does include all the languages covered in both datasets, although the additional languages used in Gemma 3's training data may still lead to increased knowledge transfer across languages.

\subsection{RQ2: How do different expansion techniques perform for varying query lengths?}
\label{sec:q_len}

The CLIRMatrix and mMARCO datasets contain queries and documents in markedly different formats and are constructed in distinct ways. Consequently, differences in the effectiveness of individual approaches across the two datasets are to be expected. In CLIRMatrix, queries correspond to Wikipedia article titles and are therefore very short, typically one or two words, while the associated documents are comparatively long. By contrast, mMARCO consists of longer, human generated queries and shorter passages. This section examines how these differences influence the behaviour of Aya Expanse 8B by comparing retrieval performance across the two datasets. Table \ref{tab:dataset_comp} summarises the performance of each cross-lingual query expansion approach on both datasets. Results for the Gemma 3 models are reported in the Appendices \ref{app:g4b} and \ref{app:g12b}. 

\begin{table}[h!]
    \centering{\includegraphics[width=0.9\linewidth]{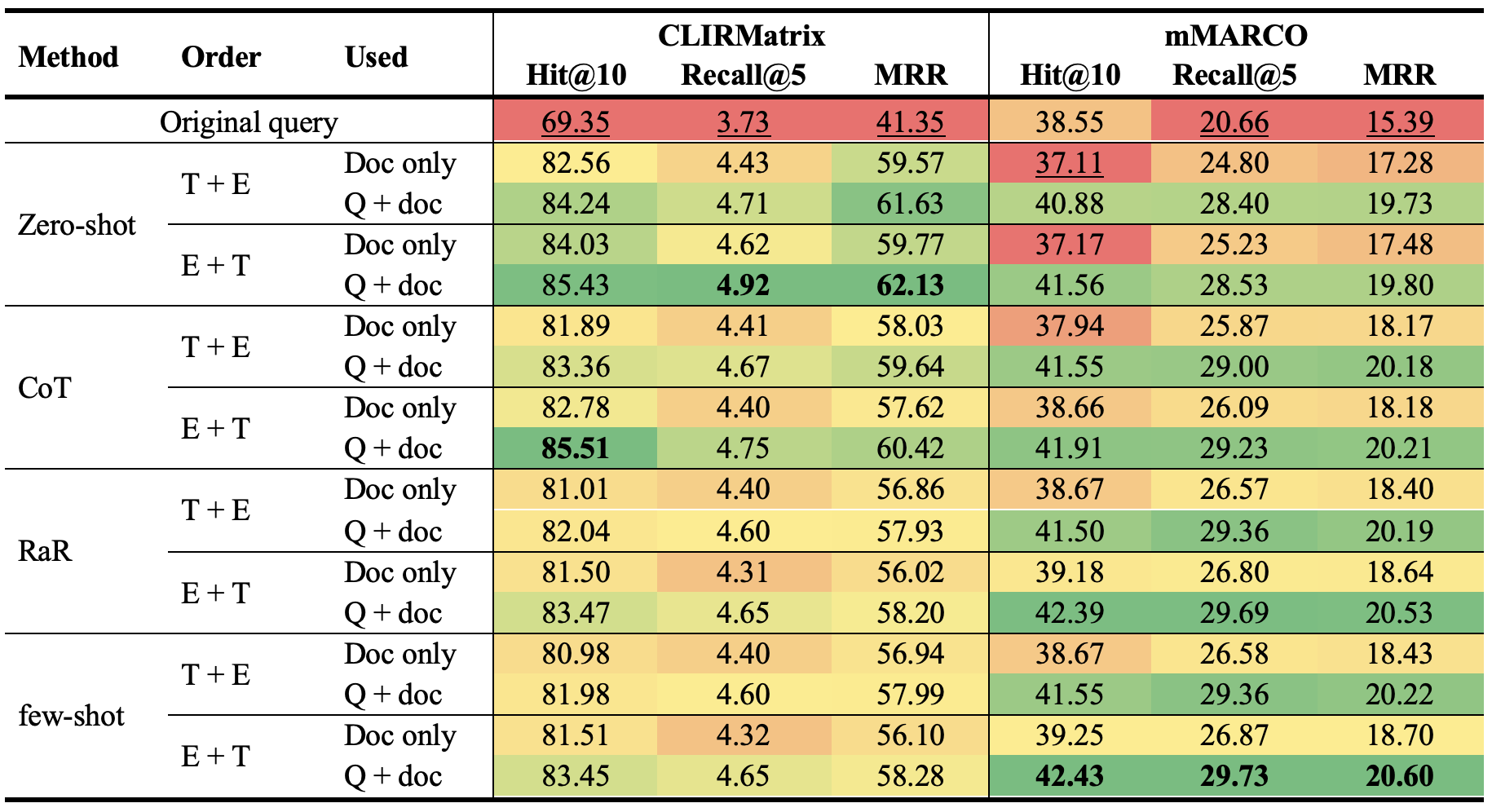}}
    \caption{Comparison of cross-lingual query expansion approaches on CLIRMatrix and mMARCO. Note that CLIRMatrix contains 100 relevant documents per query whereas mMARCO contains one per query (so Hit@k = Recall@k). (\textbf{max value}, \underline{min value})}
    \label{tab:dataset_comp}
\end{table}

For CLIRMatrix, the most pronounced differences arise between expansion techniques. In most configurations, zero-shot prompting yields retrieval performance that exceeds that of other query expansion methods, with the exception of CoT prompting when both the query and the pseudo-document are used for retrieval. This pattern can be better understood by considering the format of the CLIRMatrix queries and documents, together with qualitative inspection of expanded queries. Because the queries are extremely short, the introduction of terms and phrases that are not directly relevant can substantially degrade retrieval. For CoT prompting and RaR, the model frequently generates meta text such as \textit{``To answer this query, I will provide information about…"}, which adversely affects BM25 retrieval. In the case of few-shot prompting, the main issue arises from the examples provided to the model. Given the short queries and much longer documents, the model tends to generate content that reflects not only the target query but also the examples, leading to concept drift as irrelevant information is incorporated into the expanded query. 

A different trend emerges for mMARCO, where we can observe large variations between the alternating rows. Here, the biggest gains are obtained by concatenating the original query and pseudo-document rather than using only the pseudo-document for retrieval. Again, query length offers a plausible explanation. In CLIRMatrix, queries are so short that they are virtually always fully contained in the pseudo-document, so explicitly appending the query contributes little. Prior work has explored appending the query multiple times to partially compensate for the length imbalance between query and document \cite{wang_2023_query2doc}, which may produce a more noticeable effect for sparse retrieval. In mMARCO, however, the queries are longer, so concatenating the query and pseudo-document produces a more substantial change in the representation and a more marked effect on retrieval performance.

Consistent with the analysis in the previous subsection, the optimal expansion technique differs across datasets. Zero-shot prompting achieves the highest scores on most metrics for CLIRMatrix, whereas few-shot prompting is most effective for mMARCO. Nevertheless, in both settings, expanding the query before translation and concatenating the translated query with the pseudo-document at retrieval time generally improves performance relative to the original query. Cross-lingual query expansion yields gains of up to approximately 15\% on CLIRMatrix and 10\% on mMARCO, while differences between individual expansion techniques are smaller in magnitude.

Overall, for short title-style queries, zero-shot prompting produces the most effective pseudo-documents for retrieval due to the introduction of irrelevant information with more elaborate prompting that harms sparse retrieval. Instead, longer natural-language queries benefit from few-shot prompting, suggesting that optimal prompt design must be tailored to the length and structure of the input query. Both pre-translation expansion and concatenating the query and pseudo-document to form the expanded query are found to be most effective for higher recall.

\subsection{RQ3: How do results vary across languages?}
\label{sec:langs}

\begin{figure}[t]
    \centering{\includegraphics[width=0.65\linewidth]{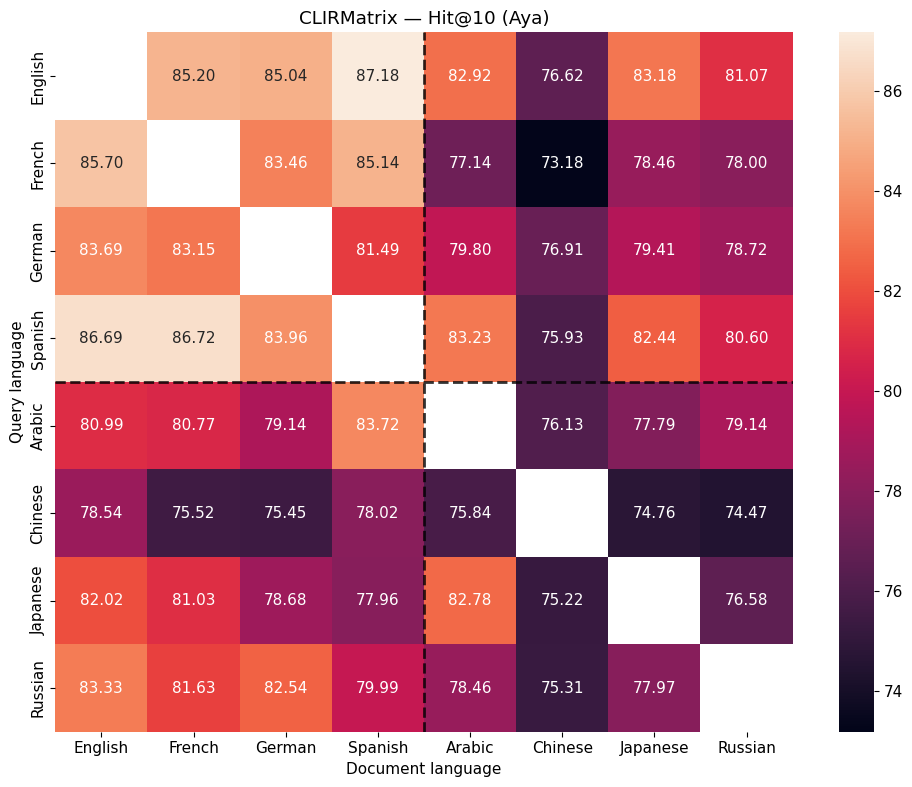}}
    \caption{Retrieval performance on CLIRMatrix across languages using cross-lingual query expansion, measured by Hit@10.}
    \label{fig:langs_cm}
\end{figure}

\begin{figure}[h]
    \centering{\includegraphics[width=0.75\linewidth]{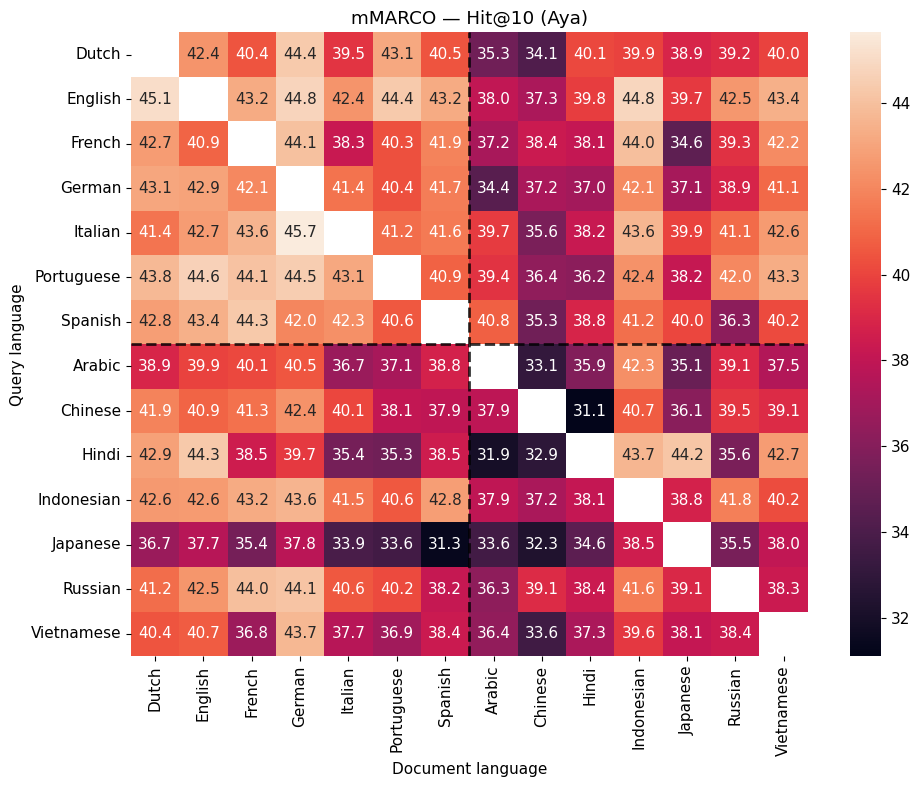}}
    \caption{Retrieval performance on mMARCO across languages using cross-lingual query expansion, measured by Hit@10.}
    \label{fig:langs_mm}
\end{figure}

The CLIRMatrix and mMARCO datasets cover different sets of languages, with mMARCO providing broader overall coverage. Retrieval performance for each language pair, measured by Hit@10, is shown in Figures \ref{fig:langs_cm} and \ref{fig:langs_mm} for CLIRMatrix and mMARCO respectively. Hit@10 indicates whether at least one relevant document appears among the top ten retrieved results for a given query. Because both datasets include several European languages that use the Latin script as well as non-European languages, the figures are divided into four quadrants, where, for example, the upper right quadrant corresponds to retrieval using queries in European languages and documents in non-European languages.

Language pairs in the upper left quadrant, which represent retrieval from one European language to another, achieve higher scores on average than those in the other three quadrants, indicating generally better performance for these combinations. In contrast, some of the remaining languages exhibit a marked drop in Hit@10. For CLIRMatrix, Chinese queries consistently yield some of the lowest values, and documents in Chinese are also among the most difficult to retrieve  (Figure \ref{fig:langs_cm}). For mMARCO, Japanese queries show particularly low performance, while documents in Arabic, Chinese and Hindi tend to be hardest to retrieve on average (Figure \ref{fig:langs_mm}).

\begin{figure}[h!]
    \centering
    \includegraphics[width=1\linewidth]{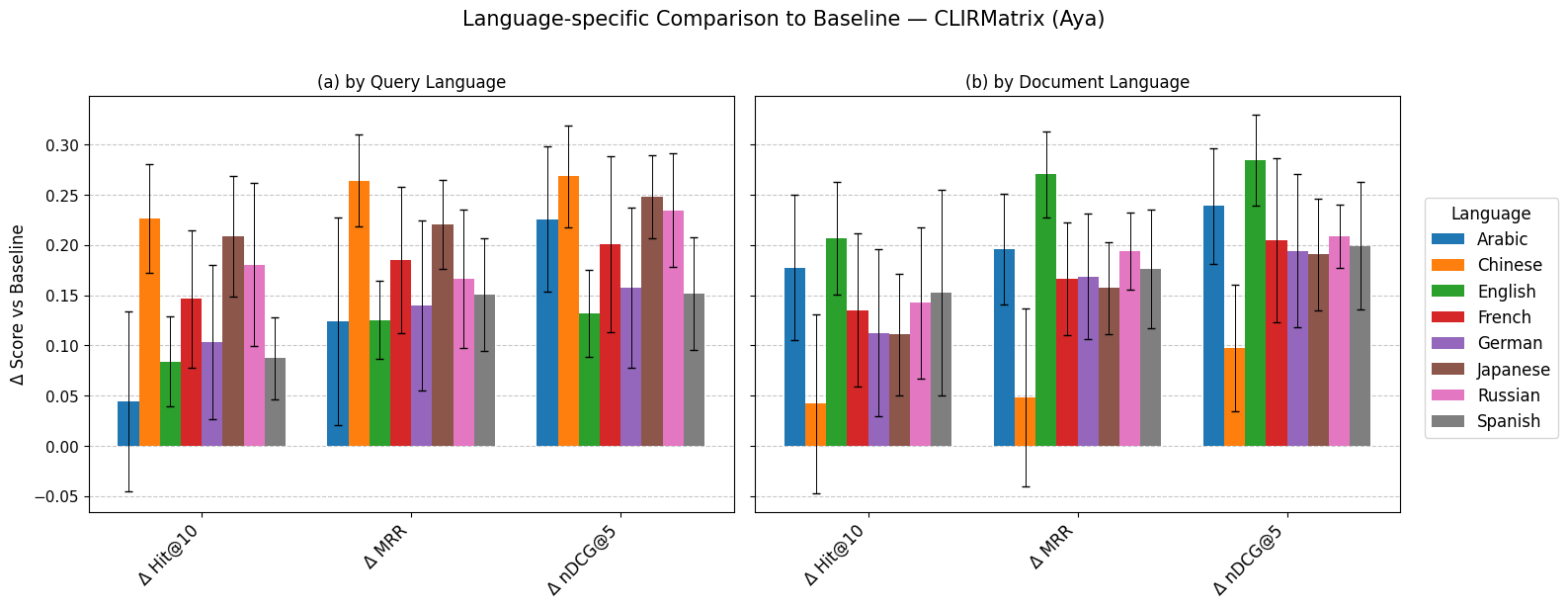}
    \caption{Cross-lingual query expansion results with Aya Expanse 8B for CLIRMatrix, change from original baseline query.}
    \label{fig:langs_change}
\end{figure}

It is relevant to consider not only on absolute performance but also on improvements from the baseline and how these vary across languages. Figure \ref{fig:langs_change} reports the change in several metrics for CLIRMatrix when applying cross-lingual query expansion, relative to retrieval using the original queries. Figure \ref{fig:langs_change}(a) aggregates results by query language and thus reflects how effective queries in a given language are at retrieving relevant documents. Figure \ref{fig:langs_change}(b) aggregates by document language and indicates how difficult it is to retrieve documents written in that language.

For CLIRMatrix, Chinese exhibits the lowest retrieval performance for both query and document languages. However, when examining differences from the baseline in Figure \ref{fig:langs_change}(a), Chinese queries benefit from the largest relative improvements from the use of cross-lingual query expansion. Thus, although the expanded Chinese queries still achieve comparatively low absolute scores, the gains over the baseline are substantial. In contrast, when focusing on document languages in Figure \ref{fig:langs_change}(b), the largest improvements in recall and related metrics are observed for English documents, whereas Chinese documents show the smallest gains.

All in all, in terms of variations across languages, we observe the best performance for European, same-script language pairs. This result is consistent with what is often reported in the literature, as is unsurprising as these are typically higher-resource and typologically similar languages. The same trend is observed across both datasets and across all models (for Gemma 3 4B and 12B results, see Appendices \ref{app:g4b} and \ref{app:g12b}, respectively). Cross-lingual query expansion does result in the biggest improvements for some of these lower performing languages, such as Chinese and Japanese for CLIRMatrix queries. However, we also see the smallest improvements when retrieving in these same languages (i.e. when attempting to retrieve documents in Chinese or Japanese with other language queries). Therefore, cross-lingual query expansion appears most effective in improving the retrieval effectiveness of languages that lead to low accuracy, but it does not seem to make it easier to retrieve documents in languages that are already hard to retrieve in.

\subsection{RQ4: Does supervised fine-tuning improve query expansion performance?}

The final set of experiments investigates whether fine-tuning a multilingual large language model on CLIR data can substantially improve retrieval performance. Fine-tuning is carried out on CLIRMatrix, which is specifically designed for cross-lingual retrieval, whereas mMARCO is formulated in a question answering style. To examine the effect of the fine-tuning on different languages, two configurations are considered. The first uses English and Spanish (En \& Es), two high-resource languages that share the Latin script and achieved strong performance in the above experiments. The second uses Arabic and Chinese (Ar \& Zh), typologically distant languages written in non-Latin scripts that tend to perform poorly, as exemplified in the results in the previous section. This experimental design allows for a direct comparison between the base Aya Expanse 8B model and variants fine-tuned on either English and Spanish or Arabic and Chinese.

% \begin{table}[t]
%     \centering
%     \includegraphics[width=1\linewidth]{figures/sft_table.png}
%     \caption{Comparison of cross-lingual query expansion performance on CLIRMatrix for all language pairs using Aya Expanse 8B and fine-tuned variant. (\textbf{max value}, \underline{min value})}
%     \label{tab:sft_table}
% \end{table}

Table \ref{tab:sft_table} summarises the results of the fine-tuned models relative to the baseline. There is a clear difference between the effects of fine-tuning on retrieval performance for CLIRMatrix and mMARCO, whereas we see smaller differences between the two fine-tuned model variants. With regards to CLIRMatrix, for almost all metrics, the fine-tuned variants yield higher retrieval scores than the base model. The only exception is for Recall@5 using RaR and few-shot prompting with the Ar \& Zh fine-tuned variant. The difference in retrieval performance when using each of the fine-tuned models for cross-lingual query expansion varies by metric and expansion technique. For CLIRMatrix, we see the smallest difference between the two models for zero-shot expansion, which is already the best-performing technique. The use of fine-tuning with the En \& Es datasets leads to improvements using other expansion techniques that match the base model's performance using zero-shot prompting.

\begin{table}[t]
    \centering
    \includegraphics[width=1\linewidth]{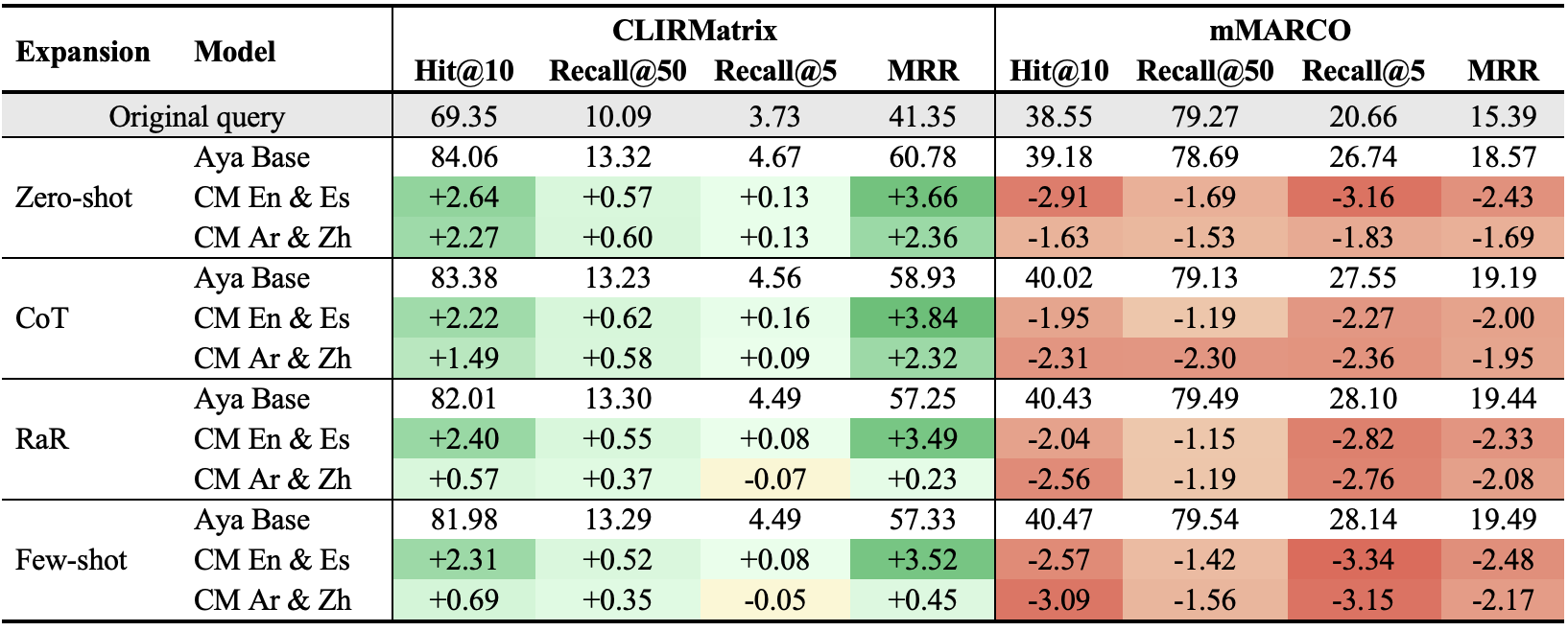}
    \caption{Comparison of cross-lingual query expansion performance for all language pairs using Aya Expanse 8B and fine-tuned variants on CLIRMatrix English and Spanish (CM En \& Es), and CLIRMatrix Arabic and Chinese (CM Ar \& Zh). For the fine-tuned models, values indicate difference from the base model. The three models shows in the table are Aya Base (Aya Expanse 8B), CM En \& Es (Aya Expanse 8B fine-tuned on the English and Spanish CLIRMatrix data), and CM Ar \& Zh (Aya Expanse 8B fine-tuned on the Arabic and Chinese CLIRMatrix data).}
    \label{tab:sft_table}
\end{table}

Instead, for mMARCO we observe a decrease in retrieval performance for every metric across both fine-tuned models as compared to the original base model. Again, there is variation in the changes from the baseline across expansion techniques and metrics. Whereas for CLIRMatrix the use of the En \& Es model for cross-lingual query expansion always leads to higher retrieval metrics than the Arabic and Chinese model, the same does not hold for retrieval on the mMARCO dataset. Notably, for zero-shot expansion, larger drops in performance result from the use of the En \& Es model than the Ar \& Zh one. 

Whereas when comparing between Aya Expanse 8B and Gemma 3 12B different expansion techniques led to more accurate retrieval, in this case there is consistency across all three models. Zero-shot expansion remains the best performing technique for the short-form queries in CLIRMatrix, whereas for the longer-form queries in mMARCO, few-shot expansion is remains persistently the best approach. Therefore, fine-tuning does not appear to alter the relative effectiveness of cross-lingual expansion strategies. Analysing the parameter combinations, previous results hold true: concatenating the query and pseudo-document outperforms retrieval with only the generated document; for translation and expansion order, neither approach leads to consistently higher metrics.

Overall, improvements through fine-tuning on multilingual information retrieval data appear domain-specific and depend on whether the intended query type to be retrieved with is similar to that which the model was fine-tuned on. Due to the difference in the type and length of queries in CLIRMatrix and mMARCO, we see that fine-tuning on the former in fact leads to losses in performance on the latter. Even for similar style queries, the gains are perhaps less marked than would have been expected, particularly as the retrieval is carried out on the same dataset that the model was fine-tuned on. However, these findings do indicate that a combination of approaches can lead to better gains in retrieval performance. Overall, generative expansion is a powerful tool for CLIR, but its effectiveness is shaped by model behaviour, query properties, and linguistic coverage.

\section{Conclusion}
\label{sec:conclusion}
%Introductory summary of the work covering aims and summary of the contributions. Future directions.

This work has analysed generative query expansion for cross-lingual information retrieval using multilingual large language models. Three open-source models (Aya Expanse 8B and Gemma 3 4B and 12B) are compared under four prompting strategies (zero-shot, Chain-of-Thought, Rephrase and Respond, and few-shot), two orders of translation and expansion, and two retrieval formulations (pseudo-document only versus query concatenated with pseudo-document) on CLIRMatrix and mMARCO. In addition, supervised fine-tuning of Aya Expanse 8B on CLIRMatrix was investigated in two language configurations (English \& Spanish, and Arabic \& Chinese).

The results indicate that LLM-based query expansion consistently improves retrieval over the original queries, although the best model and prompting strategy depend on query type. Query length emerges as a key factor: zero-shot prompting is most effective for short, title-like queries, whereas few-shot promptings is preferable for longer, natural language queries. Substantial disparities remain across languages and writing systems, with non-Latin script languages such as Chinese, Japanese, Arabic and Hindi achieving lower absolute scores despite sometimes showing the largest relative gains. Supervised fine-tuning on cross-lingual retrieval data yields consistent but modest improvements on similar-style queries, whereas it results is losses in performance when use on a different style of queries, highlighting the value of task-specific training.

These findings point to several directions for future work. There is a need for more balanced multilingual resources for cross-lingual retrieval, with broader script coverage and more diverse domains. In this work, we use a two-stage translation and expansion implementation; future research in this area may also consider zero-shot cross-lingual generation, as has been suggested for other applications \cite{wang_2023_cls, park_2025}. Beyond prompting, further study is required on how interventions such as supervised fine-tuning or reinforcement learning can strengthen query expansion \cite{li_2025_qe}. In this work, we only show the effects of fine-tuning on one dataset; future work could look at fine-tuning on mixed datasets, as this may lead to gains across domains and on different types and styles of queries. Evaluation could also move beyond downstream retrieval metrics to include direct assessment of expanded queries, for example with respect to faithfulness to the original query, concept drift, and cross-lingual consistency.

\paragraph{Limitations.}
This study has several limitations. First, BM25 is used as the retrieval algorithm to assess how much cross-lingual query expansion improves retrieval performance, following common practice in the literature. Recent work, however, has increasingly centred on dense retrieval, which relies on embeddings rather than keyword matching.  Query expansion has been shown to benefit both sparse and dense methods \cite{wang_2023_query2doc, zhang_2024_qe}, although the relative behaviour of different techniques may vary. Second, the analysis is restricted to a limited number of models and datasets. Only three open- source multilingual LLMs are considered, and commercial closed source models are excluded in order to facilitate replication of the reported results. Similarly, only two datasets with different query types and data sources are used, so the empirical scope could be broadened in future work. Finally, the original query is employed as the sole baseline when presenting results, reflecting the focus on the overall impact of cross-lingual query expansion. Isolating the individual effects of translation and expansion would require an intermediate condition that evaluates both the translated query and the expanded query in the source language. Despite these limitations, the results provide empirical evidence that generative cross-lingual query expansion is a promising component of robust and equitable cross-lingual retrieval systems.

\bibliographystyle{unsrt}  
\bibliography{main}  %%% Remove comment to use the external .bib file (using bibtex).
%% and comment out the ``thebibliography'' section.


% George Kour and Raid Saabne.
% \newblock Real-time segmentation of on-line handwritten arabic script.
% \newblock In {\em Frontiers in Handwriting Recognition (ICFHR), 2014 14th
%   International Conference on}, pages 417--422. IEEE, 2014.

% George Kour and Raid Saabne.
% \newblock Fast classification of handwritten on-line arabic characters.
% \newblock In {\em Soft Computing and Pattern Recognition (SoCPaR), 2014 6th
%   International Conference of}, pages 312--318. IEEE, 2014.

% \bibitem{hadash2018estimate}
% Guy Hadash, Einat Kermany, Boaz Carmeli, Ofer Lavi, George Kour, and Alon
% \newblock Estimate and replace: A novel approach to integrating deep neural
%   networks with existing applications.
% \newblock {\em arXiv preprint arXiv:1804.09028}, 2018.

% \end{thebibliography}

\newpage

\begin{appendices}

\section{LLM prompts}
\label{app:prompts}

System messages:
\begin{itemize}
    \item \textbf{Zero-shot, Chain-of-Thought and Rephrase and Respond}
    \begin{quote}
        You are a text expansion model. Respond only with the requested passage. Stop naturally at the end of the passage and avoid repetition.
    \end{quote}
    \item \textbf{Few-shot}
    \begin{quote}
        Given example query–passage pairs, produce one short, relevant passage about the final query. Respond only with the passage; stop naturally at the end of the passage and avoid repetition.
    \end{quote}
\end{itemize}

Prompts:
\begin{itemize}
    \item \textbf{Zero-shot} \cite{gao_2023_hyde} 
    \begin{quote}
        Please write a passage to answer the question.
            
        Question:
    \end{quote}
    \item \textbf{Chain-of-Thought} \cite{jagerman_2023}
    \begin{quote}
        Answer the following query, give the rationale before answering:
    \end{quote}
    \item \textbf{Rephrase and Respond} \cite{deng_2024_rar}
    \begin{quote}
        Rephrase and expand the question, and respond.
    \end{quote}
    \item \textbf{Few-shot} \cite{wang_2023_query2doc}
    \begin{quote}
        Please write a passage that answers the given query:

        Query: \textit{Example query 1}
        
        Passage: \textit{Example passage 1}

        Query: \textit{Example query 2}
        
        Passage: \textit{Example passage 2}
        
        Query: \textit{Example query 3}
        
        Passage: \textit{Example passage 3}
        
        Query: \textit{Example query 4}
        
        Passage: \textit{Example passage 4}
        
        Query: \textit{Example query 5}
        
        Passage: \textit{Example passage 5}
        
        Query: \textit{Relevant query}
        
        Passage:
    \end{quote}
\end{itemize}

\newpage

\section{Aya Expanse 8B}

\subsection{CLIRMatrix}

\begin{table*}[h]
\centering
\small
\begin{tabular}{lcccccccc}
\toprule
& \multicolumn{4}{c}{\textbf{By query language}}& \multicolumn{4}{c}{\textbf{By document language}}\\
\textbf{Language} &\textbf{Hit@10} & \textbf{Recall@}5& \textbf{MRR} & \textbf{nDCG@10}   & \textbf{Hit@10} & \textbf{Recall@5}&\textbf{MRR} &\textbf{nDCG@10}  \\
\midrule
Arabic&79.67& 3.06& 52.88& 53.36
& 80.02& 6.46&55.62&56.40
\\
Chinese& 76.09& 2.55& 50.67& 53.08
& 75.62& 4.88&49.75&51.08
\\
 English&83.03& 7.53& 58.58& 59.99
& 82.99& 2.43&59.69&60.42
\\
 French&80.15&  4.79&  55.82& 57.10
& 82.00& 3.68&57.07&58.01
\\
 German&80.45&  4.67&  54.82& 56.29
& 81.18& 3.73&56.89&57.91
\\
Japanese&79.18&  3.46&  53.98& 55.11
& 79.15& 5.59&54.28&55.46
\\
 Russian& 79.89&  4.63&  56.59& 56.59
& 78.37& 3.73&50.96&53.26
\\
 Spanish& 82.80& 4.40& 57.68& 58.91& 81.93& 4.61&56.74&57.89\\
  \bottomrule
\end{tabular}
\caption{Cross-lingual query expansion results with Aya Expanse 8B for CLIRMatrix for all query and document languages.}
\label{tab:method_clir_aya_source}
\end{table*}

\subsection{mMARCO}

\begin{table*}[h]
\centering
\small
\begin{tabular}{lcccccc}
\toprule
& \multicolumn{3}{c}{\textbf{By query language}}& \multicolumn{3}{c}{\textbf{By document language}}\\
\textbf{Language} &\textbf{Hit@10} & \textbf{Hit@5}  & \textbf{MRR}  &  \textbf{Hit@10} &\textbf{Hit@5}  &\textbf{MRR} \\
\midrule
Arabic&38.08& 25.42& 17.75
&  36.84&23.93&16.73
\\
Chinese& 39.00&  25.72& 17.97
&  35.58&22.71&16.58
\\
 Dutch&39.83&  26.99& 19.33
&  41.81&26.42&19.02
\\
 English&42.20&   28.69&  20.31
&  41.98&28.89&19.76
\\
 French&40.16&   26.65&  18.60
&  41.30&26.75&18.84
\\
German&39.95&   26.65&  18.57
&  42.86&28.60&20.11
\\
 Hindi& 38.88&   23.19&  16.45
&  37.20&23.13&16.01
\\
 Indonesian& 40.84& 27.21& 19.10
&  41.89&27.87&19.27
\\
 Italian& 41.30& 27.38& 18.55
&  39.46&26.43&18.46
\\
 Japanese& 35.32& 22.11& 16.16
&  38.45&26.10&18.37
\\
 Portuguese& 41.45& 27.93& 19.35
&  39.37&26.48&18.53
\\
 Russian& 40.27& 26.77& 18.73
&  39.18&25.88&18.66
\\
 Spanish& 40.62& 27.03& 19.03
&  39.66&27.47&18.56
\\
 Vietnamese& 38.31& 25.60& 17.90&  40.67&26.65&18.92\\
   \bottomrule
\end{tabular}
\caption{Cross-lingual query expansion results with Aya Expanse 8B for mMARCO for all query and document languages.}
\label{tab:method_results}
\end{table*}

\newpage

\section{Gemma 3 4B}
\label{app:g4b}

\begin{table}[h]
    \centering
    \begin{tabular}{lllcccccc}
    \toprule
          \multirow{2}{*}{\textbf{Method}}&  \multirow{2}{*}{\textbf{Order}}&  \multirow{2}{*}{\textbf{Used}}&  \multicolumn{3}{c}{\textbf{CLIRMatrix}}&  \multicolumn{3}{c}{\textbf{mMARCO}}\\
         &  &  &  \textbf{Hit@10}&  \textbf{Recall@50}&  \textbf{MRR}&  \textbf{Hit@10}&  \textbf{Recall@50}& \textbf{MRR}\\
         \midrule
         \multicolumn{3}{c}{Original query}&  69.35&  10.09&  41.35&  38.55&  79.27& 15.39\\
         \midrule
         \multirow{4}{*}{Zero-shot}& \multirow{2}{*}{T + E}&  Doc only&  75.14&  10.97&  49.22
&  37.33&  77.30& 17.90
\\
         &  &  Q + doc&  77.04&  12.14&  51.70
&  41.50&  79.49& 20.38
\\
         &  \multirow{2}{*}{E + T}&  Doc only&  76.62&  11.16&  50.10
&  36.25&  76.55& 17.23
\\
         &  &  Q + doc&  79.11&  12.43&  53.08&  41.03&  79.58& 19.92\\
         \midrule
         \multirow{4}{*}{CoT}&  \multirow{2}{*}{T + E}&  Doc only&  75.52&  11.04&  49.60
&  38.26&  77.02& 18.44
\\
         &  &  Q + doc&  77.82&  12.16&  51.97
&  41.68&  79.76& 20.42
\\
         &  \multirow{2}{*}{E + T}&  Doc only&  77.35&  11.40&  51.05
&  36.34&  77.30& 17.61
\\
 & & Q + doc& 79.87& 12.59& 53.91
& 40.72& 79.99&20.05\\
\midrule
 \multirow{4}{*}{RaR}& \multirow{2}{*}{T + E}& Doc only& 69.40& 10.16& 43.89
& 36.50& 76.41&17.58
\\
 & & Q + doc& 72.23& 11.28& 
46.43
& 39.83& 79.07&19.48
\\
 & \multirow{2}{*}{E + T}& Doc only& 72.44& 10.27& 
46.19
& 36.66& 77.11&17.62
\\
 & & Q + doc& 75.40& 11.62& 48.92& 40.55& 79.47&19.83\\
 \midrule
 \multirow{4}{*}{Few-shot}& \multirow{2}{*}{T + E}& Doc only& 69.47& 10.14& 43.98
& 36.50& 76.46&17.60
\\
 & & Q + doc& 72.20& 11.26& 46.48
& 39.84& 79.09&19.50
\\
 & \multirow{2}{*}{E + T}& Doc only& 72.37& 10.24& 46.24
& 36.68& 77.08&17.67
\\
 & & Q + doc& 75.29& 11.58& 48.91& 40.58& 79.51&19.88\\
 \bottomrule
    \end{tabular}
    \caption{Cross-lingual query expansion results with Gemma 3 4B for CLIRMatrix and mMARCO using different expansion techniques.}
    \label{tab:g4_methods}
\end{table}

\subsection{CLIRMatrix}

\begin{table}[h]
\centering
\small
\begin{tabular}{lcccccccc}
\toprule
& \multicolumn{4}{c}{\textbf{By query language}}& \multicolumn{4}{c}{\textbf{By document language}}\\
\textbf{Language} &\textbf{Hit@10} & \textbf{Recall@}5& \textbf{MRR} & \textbf{nDCG@10}   & \textbf{Hit@10} & \textbf{Recall@5}&\textbf{MRR} &\textbf{nDCG@10}  \\
\midrule
Arabic&72.74& 2.33& 44.65& 45.91
& 71.10& 4.98&45.25&46.90
\\
Chinese& 70.37& 2.12& 43.60& 46.64
& 68.77& 3.98&42.83&44.41
\\
 English&76.81& 6.28& 50.52& 52.72
& 78.48& 2.11&52.44&54.20
\\
 French&73.75&  3.97&  47.94& 49.92
& 76.32& 3.09&49.70&51.35
\\
 German&73.12&  3.76&  46.49& 48.56
& 74.76& 3.13&48.89&50.65
\\
Japanese&73.05&  2.96&  47.37& 48.78
& 72.92& 4.52&46.70&48.49
\\
 Russian& 73.34&  3.74&  48.27& 49.07
& 70.70& 3.00&42.92&45.75
\\
 Spanish& 76.69& 3.60& 49.99& 52.02& 76.81& 3.96&50.10&51.86\\
  \bottomrule
\end{tabular}
\caption{Cross-lingual query expansion results with Gemma 3 4B for CLIRMatrix for all query and document languages.}
\label{tab:method_clir_aya_source}
\end{table}

\begin{figure}[h]
    \centering{\includegraphics[width=0.75\linewidth]{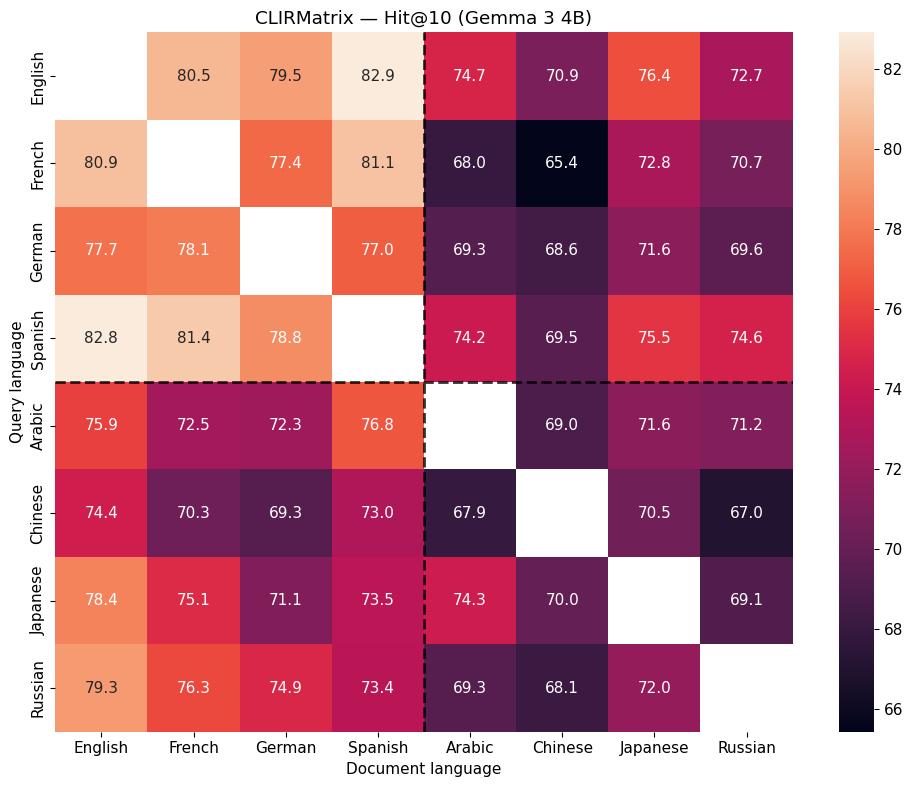}}
    \caption{Retrieval performance on CLIRMatrix across languages using cross-lingual query expansion with Gemma 3 4B, measured by Hit@10.}
    \label{fig:langs_cm_4}
\end{figure}

\newpage

\subsection{mMARCO}

\begin{table}[h]
\centering
\small
\begin{tabular}{lcccccc}
\toprule
& \multicolumn{3}{c}{\textbf{By query language}}& \multicolumn{3}{c}{\textbf{By document language}}\\
\textbf{Language} &\textbf{Hit@10} & \textbf{Hit@5}  & \textbf{MRR}  &  \textbf{Hit@10} &\textbf{Hit@5}  &\textbf{MRR} \\
\midrule
Arabic&37.95& 25.56& 18.39
&  35.59&23.04&16.26
\\
Chinese& 37.47&  24.40& 17.24
&  35.39&22.44&16.21
\\
 Dutch&38.49&  25.87& 18.75
&  41.85&26.68&19.54
\\
 English&41.88&   28.41&  20.22
&  40.93&27.69&19.29
\\
 French&39.55&   26.72&  18.92
&  40.16&25.97&18.90
\\
German&39.93&   26.57&  18.77
&  41.58&27.38&19.26
\\
 Hindi& 37.78&   22.90&  16.40
&  37.09&23.13&16.03
\\
 Indonesian& 39.70& 25.95& 18.11
&  39.61&26.16&18.74
\\
 Italian& 39.83& 25.73& 18.05
&  39.04&26.29&18.60
\\
 Japanese& 34.33& 20.99& 15.96
&  36.42&23.85&17.01
\\
 Portuguese& 39.39& 26.24& 18.81
&  38.00&25.75&18.36
\\
 Russian& 39.11& 25.67& 18.13
&  38.87&25.37&18.85
\\
 Spanish& 39.03& 25.81& 18.47
&  38.85&26.56&18.41
\\
 Vietnamese& 37.69& 25.06& 17.64&  38.76&25.57&18.39\\
   \bottomrule
\end{tabular}
\caption{Cross-lingual query expansion results with Gemma 3 4B for mMARCO for all query and document languages.}
\label{tab:method_results}
\end{table}

\newpage

\begin{figure}[h]
    \centering{\includegraphics[width=0.75\linewidth]{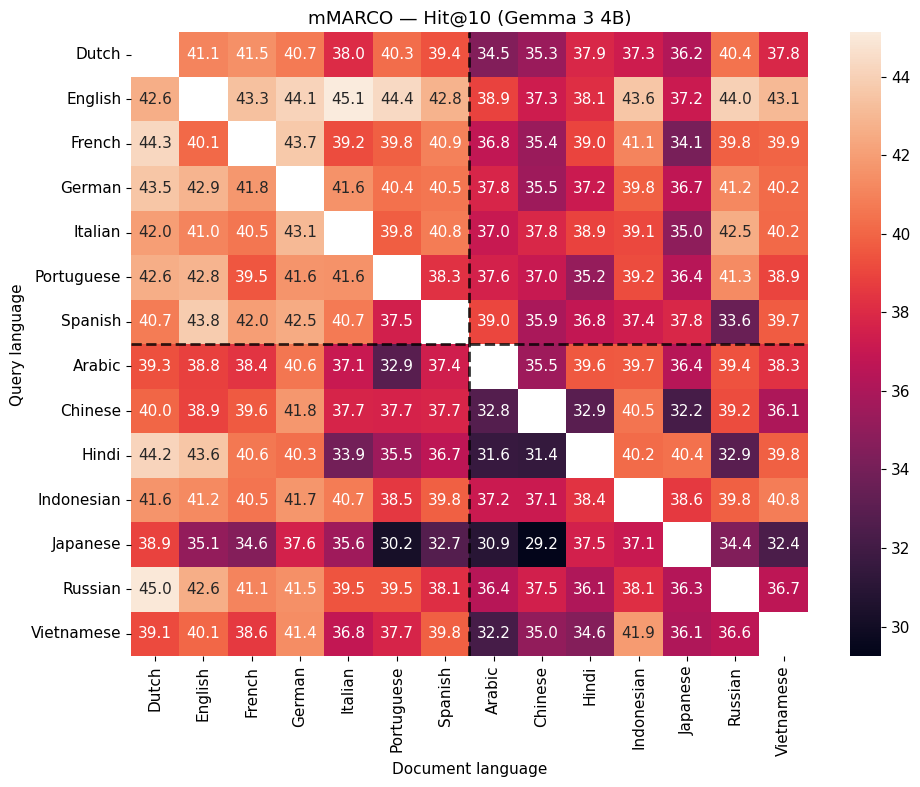}}
    \caption{Retrieval performance on mAMRCO across languages using cross-lingual query expansion with Gemma 3 4B, measured by Hit@10.}
    \label{fig:langs_mm_4}
\end{figure}

\newpage

\section{Gemma 3 12B}
\label{app:g12b}

\begin{table}[h]
    \centering
    \begin{tabular}{lllcccccc}
    \toprule
          \multirow{2}{*}{\textbf{Method}}&  \multirow{2}{*}{\textbf{Order}}&  \multirow{2}{*}{\textbf{Used}}&  \multicolumn{3}{c}{\textbf{CLIRMatrix}}&  \multicolumn{3}{c}{\textbf{mMARCO}}\\
         &  &  &  \textbf{Hit@10}&  \textbf{Recall@50}&  \textbf{MRR}&  \textbf{Hit@10}&  \textbf{Recall@50}& \textbf{MRR}\\
         \midrule
         \multicolumn{3}{c}{Original query}&  69.35&  10.09&  41.35&  38.55&  79.27& 15.39\\
         \midrule
         \multirow{4}{*}{Zero-shot}& \multirow{2}{*}{T + E}&  Doc only&  83.64&  12.57&  60.48
&  38.34&  78.12& 18.81
\\
         &  &  Q + doc&  84.87&  13.78&  62.26
&  42.47&  80.44& 20.96
\\
         &  \multirow{2}{*}{E + T}&  Doc only&  85.26&  12.67&  61.49
&  39.05&  78.35& 18.78
\\
         &  &  Q + doc&  86.69&  13.96&  63.89&  42.69&  80.77& 21.11\\
         \midrule
         \multirow{4}{*}{CoT}&  \multirow{2}{*}{T + E}&  Doc only&  79.93&  12.03&  57.09
&  39.88&  79.20& 19.16
\\
         &  &  Q + doc&  81.93&  13.23&  58.84
&  43.23&  81.36& 21.53
\\
         &  \multirow{2}{*}{E + T}&  Doc only&  83.87&  12.75&  59.91
&  39.83&  79.19& 19.27
\\
 & & Q + doc& 85.06& 13.87& 61.90& 43.65& 81.49&21.56\\
\midrule
 \multirow{4}{*}{RaR}& \multirow{2}{*}{T + E}& Doc only& 77.46& 12.07& 53.76
& 39.63& 79.07&19.00
\\
 & & Q + doc& 79.41& 13.30& 
55.28
& 42.46& 81.54&20.67
\\
 & \multirow{2}{*}{E + T}& Doc only& 79.34& 12.03& 
54.43
& 40.78& 80.20&19.75
\\
 & & Q + doc& 82.15& 13.42& 56.84& 43.82& 82.21&21.45\\
 \midrule
 \multirow{4}{*}{Few-shot}& \multirow{2}{*}{T + E}& Doc only& 77.46& 12.05& 53.82
& 39.69& 79.09&19.05
\\
 & & Q + doc& 79.36& 13.28& 55.30
& 42.46& 81.55&20.71
\\
 & \multirow{2}{*}{E + T}& Doc only& 79.38& 12.01& 54.48
& 40.83& 80.23&19.78
\\
 & & Q + doc& 82.18& 13.40& 56.86& 43.83& 82.21&21.49\\
 \bottomrule
    \end{tabular}
    \caption{Cross-lingual query expansion results with Gemma 3 12B for CLIRMatrix and mMARCO using different expansion techniques.}
    \label{tab:g12_methods}
\end{table}

\newpage

\subsection{CLIRMatrix}

\begin{table}[h]
\centering
\small
\begin{tabular}{lcccccccc}
\toprule
& \multicolumn{4}{c}{\textbf{By query language}}& \multicolumn{4}{c}{\textbf{By document language}}\\
\textbf{Language} &\textbf{Hit@10} & \textbf{Recall@}5& \textbf{MRR} & \textbf{nDCG@10}   & \textbf{Hit@10} & \textbf{Recall@5}&\textbf{MRR} &\textbf{nDCG@10}  \\
\midrule
Arabic&78.28& 2.98& 51.83& 52.42
& 76.77& 5.94&51.90&53.07
\\
Chinese& 75.66& 2.50& 50.60& 53.03
& 74.09& 4.76&48.68&49.95
\\
 English&82.34& 7.45& 58.49& 59.70
& 83.11& 2.43&59.76&60.72
\\
 French&79.72&  4.74&  54.86& 56.32
& 81.30& 3.64&56.51&57.42
\\
 German&80.49&  4.68&  55.44& 56.81
& 81.62& 3.79&57.99&58.79
\\
Japanese&77.71&  3.36&  53.22& 54.19
& 78.87& 5.62&54.93&55.77
\\
 Russian& 78.86&  4.55&  55.79& 55.79
& 76.37& 3.67&50.06&52.25
\\
 Spanish& 81.09& 4.19& 56.57& 57.83& 82.03& 4.59&56.98&58.10\\
  \bottomrule
\end{tabular}
\caption{Cross-lingual query expansion results with Gemma 3 12B for CLIRMatrix for all query and document languages.}
\label{tab:method_clir_aya_source}
\end{table}

\begin{figure}[h]
    \centering{\includegraphics[width=0.75\linewidth]{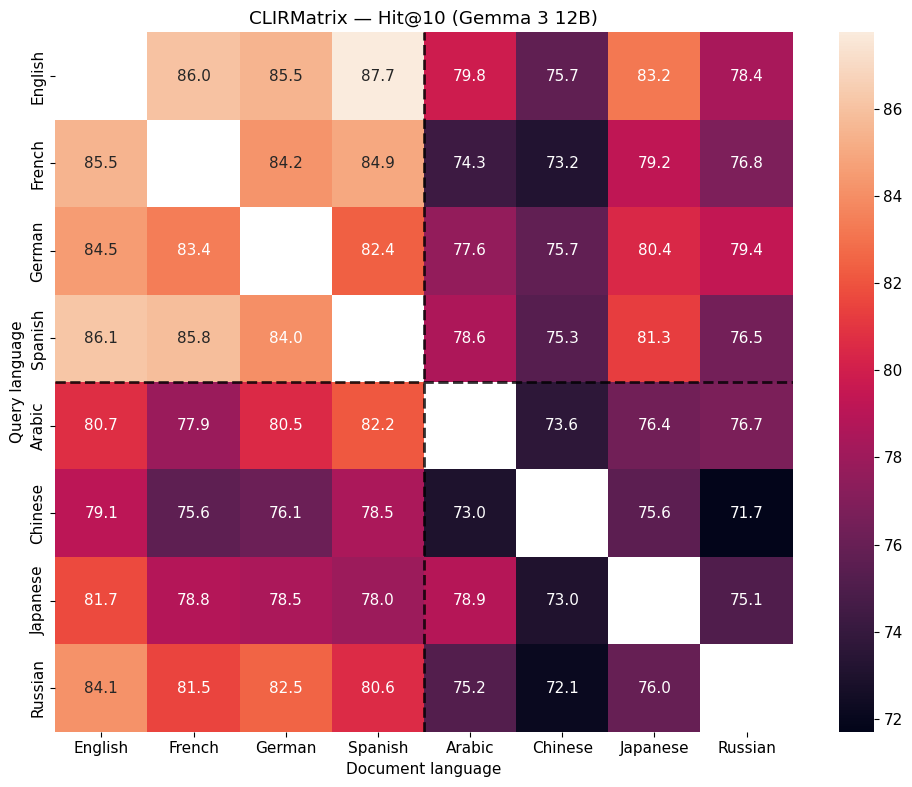}}
    \caption{Retrieval performance on CLIRMatrix across languages using cross-lingual query expansion with Gemma 3 12B, measured by Hit@10.}
    \label{fig:langs_cm_12}
\end{figure}

\newpage

\subsection{mMARCO}

\begin{table}[h]
\centering
\small
\begin{tabular}{lcccccc}
\toprule
& \multicolumn{3}{c}{\textbf{By query language}}& \multicolumn{3}{c}{\textbf{By document language}}\\
\textbf{Language} &\textbf{Hit@10} & \textbf{Hit@5}  & \textbf{MRR}  &  \textbf{Hit@10} &\textbf{Hit@5}  &\textbf{MRR} \\
\midrule
Arabic&39.33& 26.90& 18.80
&  36.19&23.59&16.86
\\
Chinese& 40.80&  26.85& 18.93
&  37.99&24.46&17.50
\\
 Dutch&41.00&  27.98& 19.96
&  43.47&28.49&20.40
\\
 English&44.01&   30.24&  21.43
&  43.32&29.79&20.83
\\
 French&41.49&   27.68&  19.52
&  42.60&27.89&19.51
\\
German&40.65&   27.40&  19.46
&  43.86&28.40&19.58
\\
 Hindi& 39.59&   24.47&  17.79
&  39.54&25.39&17.64
\\
 Indonesian& 41.84& 27.94& 19.37
&  42.01&28.85&20.38
\\
 Italian& 41.76& 27.52& 18.99
&  40.68&28.14&19.74
\\
 Japanese& 37.34& 23.84& 17.65
&  38.67&25.44&18.28
\\
 Portuguese& 41.84& 28.11& 19.83
&  40.83&27.81&19.51
\\
 Russian& 41.18& 27.42& 19.43
&  40.87&27.25&20.13
\\
 Spanish& 42.22& 28.47& 19.72
&  41.89&28.70&19.75
\\
 Vietnamese& 38.72& 26.02& 18.35&  39.88&26.64&19.12\\
   \bottomrule
\end{tabular}
\caption{Cross-lingual query expansion results with Gemma 3 12B for mMARCO for all query and document languages.}
\label{tab:method_results}
\end{table}

\begin{figure}[h]
    \centering{\includegraphics[width=0.75\linewidth]{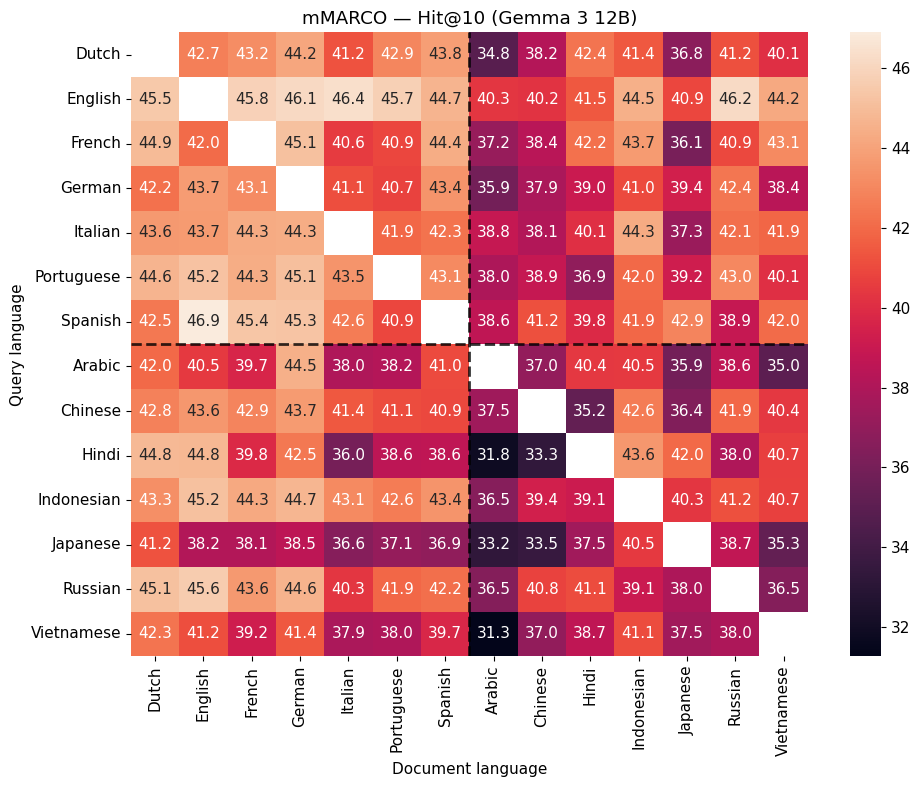}}
    \caption{Retrieval performance on mMARCO across languages using cross-lingual query expansion with Gemma 3 12B, measured by Hit@10.}
    \label{fig:langs_mm_12}
\end{figure}

\end{appendices}

\end{document}